\documentclass[11pt]{article}
\usepackage[utf8x]{inputenc}
\usepackage{graphicx,subfigure,color}
\usepackage{latexsym,amssymb,amsmath,amsfonts,amsthm,enumerate}
\usepackage{amscd}
\usepackage{wasysym}
\usepackage[T1]{fontenc}
\usepackage{authblk}

\begin{document}

\title{}

\title{Mutual information and Bose-Einstein condensation}
\author[1]{C. N. Gagatsos\thanks{cgagatso@ulb.ac.be}}
\author[2]{A. I. Karanikas\thanks{akaran@phys.uoa.gr}}
\author[2,3]{G. I. Kordas\thanks{g.kordas@thphys.uni-heidelberg.de}}
\affil[1]{\textit{Quantum Information and Communication, Ecole Polytechnique de Bruxelles,
Universit\'e Libre de Bruxelles, 1050 Brussels, Belgium}}
\affil[2]{\textit{University of Athens, Physics Department, Nuclear \& Particle Physics\\
Section Panepistimiopolis, Ilissia GR-15771, Athens, Greece}}
\affil[3]{\textit{Institut f\"ur theoretische Physik and Center for Quantum\\
Dynamics, Universit\"at Heidelberg, D--69120 Heidelberg, Germany}}
\date{}
\maketitle

\begin{abstract}
In the present work we are studying a bosonic quantum field system at finite temperature, and at zero and non-zero chemical potential.
For a  simple spatial partition we derive the corresponding mutual information, a quantity that measures the total amount of information 
of one of the parts about the other. In order to find it, we first derive the geometric entropy  corresponding to the specific partition 
and then we substract its extensive part which coincides with the thermal entropy of the system. In the case of non-zero chemical potential,
we examine the influence of the underlying Bose-Einstein condensation on the behavior of the mutual information, and we find that its 
thermal derivative possesses a finite discontinuity at exactly the critical temperature. 
\end{abstract}


\newpage
\section{Introduction}
The notion of the geometric entropy, in the framework of a quantum system, is quite old. One of the first calculations \cite{Bomb86}
was performed 
in the eighties for the case of a scalar field propagating in a black hole background. Some years later, a similar problem, in the 
framework of a quantum field theory, was addressed by several authors \cite{Sred93,Kabat94,Callan94,Calabrese04}.  Geometric entropy, generally speaking, is a measure of 
the information loss after cutting out a spatial region of the system. It caught attention because of its characteristic behavior: 
for a system in its ground state, it grows like the boundary surface of the excluded subregion, a property it shares with the 
black hole entropy.  In fact, in the context of quantum field theory, pioneering work on the geometric entropy was driven in part 
by the suggested connection to the Bekenstein –Hawking black hole entropy \cite{Bekenstein73}. From the very beginning, geometric entropy
has been tightly related with the presence of spatial entanglement in a quantum system. Entanglement 
is a fundamental ingredient of quantum mechanics leading to strong correlations between subsystems.
From the early days of quantum mechanics up until now, it has been playing an increasingly important 
role in understanding and controlling  quantum systems.  The interest in it has been renewed \cite{Plenio98} after the 
developments of the quantum information science in which it is viewed as a resource in quantum information processing.
Geometric entropy has been considered as a measure of spatial entanglement when the system under consideration is in a
pure quantum state with a density matrix of the form $\hat{\rho}=|\Psi \rangle \langle \Psi|$. By defining an “in” and an “out”
spatial region and tracing out the
“in” degrees of freedom one obtains the reduced density matrix for the “out” region: $\hat{\rho}_{\rm{out}}=\rm{tr}_{in}\hat{\rho}$.
The geometric entropy is then defined as the von Neumann entropy: $S_{\rm{out}}=-\textrm{tr} \hat{\rho}_{\rm{out}} \ln \hat{\rho}_{\rm{out}}$.

When the system is in a thermal state, e.g. in a mixed state, the geometric entropy can be defined, following the von Neumann definition,
in an analogous way. However, in this case, it does not have the same properties as the entanglement entropy in a pure state system, and it
is no longer a good estimator of entanglement since it mixes correlations of different types \cite{Callan94,Calabrese04,Wolf08},
from genuine quantum to
thermal correlations. Since it measures the thermal information loss, geometric entropy becomes an extensive quantity at the
limit of an infinite system, and loses the “area law” behavior that characterizes a pure state system. As an alternative probe for
the amount of correlations between different parts of a system in the case of thermal states,the notion of the so-called “mutual information” \cite{Adami97} has been proposed, which,
roughly speaking, eliminates the contribution of the extensive part of the thermal entropy
from the geometric entropy and can be considered as an upper limit for the entanglement entropy.

In any case, geometric entropy has been considered as a convenient construction, playing the role of an order parameter,
for the investigation of finite temperature conformal quantum field systems \cite{Calabrese04} and in the context of the AdS/CFT correspondence, aiming
at the physics of strongly coupled Quark-Gluon Plasma, the weakly coupled deconfined phase of Yang-Mills theories or the phase structure
of large $N$ QCD at a finite density \cite{Fujita08}. 

In the current work, we examine the geometric entropy in a free bosonic quantum field theory at finite temperature and at zero and
non-zero chemical potential.  Having found it, we subtract its extensive part, that is, the part related to the amount of information
that is lost due to the mixed nature of the system. The result, defined as the mutual information, quantifies the spatial correlations
between the different parts of the system and exhibits the known area law behavior.

The underlying reason for  the present study is connected to the Bose-Einstein condensation that characterizes the system, which has been
in the center of theoretical and experimental investigations during the last fifteen years after the production of the condensate in
the laboratory \cite{Anderson95}. Bose-Einstein condensation has the characteristics of a phase transition albeit, theoretically at least, it can
take place in an ideal system \cite{Landsberg65,Kapusta81,Haber81,Pitaevskii03}. It is then natural to search for the interconnection between 
this phase transition and the spatial correlations
 in the Bose system. Our findings indicate that, indeed, the Bose-Einstein condensation influences the behavior of the mutual
information: We find that its derivative with respect to the temperature, $\partial{I_m}/\partial \textrm{T}$, has a finite discontinuity
at the critical temperature both at the non-relativistic and the relativistic limit. Thus, we show how this phase 
transition leaves its fingerprint on a quantum informational quantity like mutual information.

The paper is structured as follows:  In section II we present the calculation of the geometric entropy at finite temperature
and zero chemical potential and we define the resulting mutual information, setting the stage for our result goal which is the topic of the
section that follows. This is section III, in which we apply our results in an environment with a finite charge density, we calculate the
explicit form of the mutual information, and we derive the discontinuity of its temperature derivative. Finally, in the last section we
summarize our findings. In Appendix A we present some technical details of our calculations.

\section{Geometric entropy at finite temperature}
Our starting point is the thermal density matrix $\hat{\rho}=\textrm{e}^{-\beta\hat{H}}/Z$ of a quantum field system and its Fock
space representation:
\begin{eqnarray}
\rho[\Phi',\Phi]=\frac{1}{Z}\langle\Phi'|\textrm{e}^{-\beta\hat{H}}|\Phi\rangle, \label{1}
\end{eqnarray}
where $\Phi$ denotes a single scalar field or a collection of fields. The matrix element (\ref{1}) can be written as a functional integral:                           
\begin{eqnarray}
\rho[\Phi',\Phi]=\mathop{\frac{1}{Z(\beta)}\int \mathcal{D}\Phi(\tau,\vec{x})}\limits_{\substack {\Phi(0,\vec{x})=\Phi(\vec{x}) \\ \Phi(\beta,\vec{x})=\Phi'(\vec{x})}}
\rm{e}^{-\mathop{\int} \limits_{0}^{\beta} d\tau \int d^D x \mathcal{L}[\Phi]}.  \label{2}
\end{eqnarray}                                                          
It worth noting that at the zero temperature limit, $\beta\rightarrow\infty$, (\ref{2}) is  just an interpretation of the
ground state density matrix \cite{Callan94}. It is then natural to expect that our result will reproduce, at this limit, the known
\cite{Kabat94,Callan94,Calabrese04} entanglement entropy. To derive the geometric entropy we follow the usual line of reasoning and we divide the
$D$ dimensional space on which our system is defined,
into two regions $A:\ (x_1>0,\vec{x}_{\bot})$ and $B:\ (x_1<0,\vec{x}_{\bot})$. Tracing out the “in” region, and gluing along the axis $x_1>0$,
$n$ copies of the resulting reduced density matrix, we find \cite{Kabat94,Callan94,Calabrese04}:
\begin{eqnarray}
 \textrm{tr}(\hat{\rho}_R)^n=\frac{1}{Z^n(\beta)}\mathop \int \limits_{M_n} \mathcal{D}\Phi \textrm{e}^{-S[\Phi]} \equiv \frac{Z_n(\beta)}{Z^n(\beta)}.
 \label{3}
\end{eqnarray}                                                                                   
In $Z_n$ the fields are defined on a $D+1$ dimensional space $M_n=R_{D-1} \times C_n $. The subspace $R_{D-1}$ is an Euclidean space
with metric $ds^2=dx_2^2+\dots+dx_D^2$ while $C_n$ is a two
dimensional Riemann space consisting of $n$  sheets glued together along the positive $x_1$ axis.
This $n$ folded structure turns eventually \cite{Callan94} the $(\tau,x_1)$ plane into a flat cone with an angle deficit
$\delta=2 \pi (1-n)$ at the origin. 
Having found $\textrm{tr}(\hat{\rho}_R)^n$ the geometric entropy is defined through the relation:
\begin{eqnarray}
 -\lim_{n\rightarrow 1}\frac{\textrm{tr}(\hat{\rho}_R)^n-1}{n-1}&=&-\textrm{tr}(\hat{\rho}_R \ln \hat{\rho}_R )
 \equiv S_g=-\Big(\frac{\partial}{\partial n}-1\Big) \ln Z_n \Bigg|_{n=1}.
\label{4}
\end{eqnarray}                                         
For a free bosonic theory, the partition function  can be deduced by following standard steps:
\begin{eqnarray}\nonumber
 \ln Z_n (\beta)&=&\frac{1}{2} \mathop \int \limits_{0^+}^{\infty} \frac{dT}{T} \textrm{e}^{-Tm^2}\textrm{tr}_{M_n} \textrm{e}^{-T(-\partial_E^2)}
 =\frac{1}{2}\frac{1}{(4\pi)^{(D-1)/2}}V_{D-1} \times\\
 &&\times \mathop \int \limits_{0^+}^{\infty} \frac{dT}{T^{(D+1)/2}}\textrm{e}^{-Tm^2}\textrm{tr}_{C_n}
\textrm{e}^{-T(-\partial^2_E)}
\label{5}
\end{eqnarray}
where $\partial_E^2=\partial_\tau^2+\partial_{\vec{x}}^2+m^2$.
Due to the locality of the action, the partition function in (\ref{5}) is not expected to depend explicitly
on the details of the Riemann surface. Thus, in order to calculate the non-trivial trace appearing in (\ref{5}),
we start with the finite temperature propagator of a free particle in cartesian coordinates:                                     
\begin{eqnarray}
A_{\beta_n}(\vec{x}',\vec{x})=\langle \vec{x}'|\textrm{e}^{-T(-\partial_E^2)}| \vec{x} \rangle_{\beta_n}. \label{6}
\end{eqnarray}
The subscript $\beta_n=\beta n$ indicates the periodic boundary conditions imposed on the thermal Green’s function.
As it is obvious, they are dictated by the $n$ folded structure of the Riemann space $C_n$.
The next step is to transfer the result onto a two dimensional cone with angle deficit$2\pi(1-n)$: 
\begin{eqnarray}
ds^2=d\rho^2+\rho^2 n^2 d\theta^2,\ 0\leq\theta\leq 2 \pi. \label{7}
\end{eqnarray}
One can easily find that the free thermal propagator in (\ref{6}) assumes the form  \cite{Simonov95}:
\begin{eqnarray}\nonumber
 A_{\beta_n}(\vec{x}',\vec{x})
&=&\frac{1}{4\pi T} \sum_{\nu=-\infty}^{\infty} \textrm{e}^{-\frac{1}{4T}\Big[(x_1'-x_1)^2+(x_0'-x_0-\nu \beta_n)^2\Big]}\\
 &=&\frac{1}{4\pi T} \sum_{\nu=-\infty}^{\infty} \textrm{e}^{-\frac{1}{4T}(\vec{x}'-\vec{x})^2
+\frac{\nu\beta_n}{2T}(x_0'-x_0)-\frac{(\nu \beta_n)^2}{4T}}.
\label{8}
\end{eqnarray}

The above expression can be written in the conical metric (\ref{7}) by making the replacements $x_0=\rho\sin(n\theta),\ x_1=\rho\cos(n\theta)$,
and using the expansion \cite{Deser88}:
\begin{eqnarray}
\textrm{e}^{iz\cos(n\theta)}=\sum_{m=-\infty}^{\infty}c_m J_{\frac{|m|}{n}}(z)\textrm{e}^{im\theta},\ c_m=i^{\frac{|m|}{n}}, \label{9}
\end{eqnarray}
where $J_{\frac{|m|}{n}}$ are Bessel functions of the first kind.
Thus, the thermal propagator on the surface (\ref{7}) reads:
\begin{eqnarray}\nonumber
 A_{\beta_n}(\rho',\theta';\rho,\theta;n)&=&\frac{1}{4\pi i t}\sum_{\nu,m,m_1,m_2}\textrm{e}^{-\frac{(\nu \beta)^2}{4it}}
\textrm{e}^{-\frac{\rho'^2+\rho^2}{4it}}\times\\
 \nonumber&&\times\textrm{e}^{im(\theta'-\theta)}\textrm{e}^{im_1\theta'+im_2\theta}
 \textrm{e}^{-\frac{i\pi (m_1+m_2)}{2n}}J_{\frac{|m|}{n}}\Big(\frac{\rho'\rho}{2t}\Big)\times\\
  \nonumber&&\times J_{\frac{|m_1|}{n}}\Big(\frac{\nu \beta_n}{2t} \rho' \Big) J_{\frac{|m_2|}{n}}\Big(\frac{\nu \beta_n}{2t} \rho \Big)\times\\
 &&\times i^{-\frac{|m|}{n}} i^{\frac{|m_1|}{n}}(-i)^{\frac{|m_2|}{n}}.
\label{10}
\end{eqnarray}
In the last expression the rotation $T\rightarrow it$ has been adopted in order to secure convergence of all our intermediate steps. 

Tracing out (\ref{10}) we find:
\begin{eqnarray}\nonumber
 A_{\beta_n}&=&\textrm{tr}_{C_n} \textrm{e}^{-T(-\partial_E^2)}\\
&=&\frac{1}{2it}\sum_{\nu,m,m_1}\textrm{e}^{-\frac{(\nu\beta)^2}{4it}} i^{-\frac{|m|}{n}}
\mathop \int \limits_0^\infty d\rho \rho \textrm{e}^{-\frac{\rho^2}{2it}} J_{\frac{|m|}{n}}\Big(\frac{\rho^2}{2t}\Big)
J_{\frac{|m_1|}{n}}^2\Big(\frac{\nu \beta_n}{2t} \rho\Big).
\label{11}
\end{eqnarray}    
At this point we stress the fact that for $n\neq1$, the trace over the conical metric (\ref{7}), that is the integration over $\rho$,
must be performed before the summations over $m$ or $m_1$. The relevant calculations can be facilitated by using the fact that
we are only interested in the derivative of (\ref{11}) with respect to $n$:
\begin{eqnarray}\nonumber
 \big(\partial_n A_{\beta_n}\big)_{n=1}&=&
\frac{1}{2it}\sum_{\nu} \textrm{e}^{-\frac{(\nu \beta)^2}{4it}} \partial_n \Bigg[\sum_m \mathop \int \limits_0^{\infty} d\rho \rho
\textrm{e}^{-\frac{\rho^2}{2it}} i^{-\frac{|m|}{n}} J_{\frac{|m|}{n}} \Big(\frac{\rho^2}{2t}\Big)\Bigg]_{n=1}+\\
 &&+\frac{1}{2it} \partial_n \Bigg[\sum_{\nu,m}  \textrm{e}^{-\frac{(\nu \beta_n)^2}{4it}} \mathop \int \limits_0^{\infty} d\rho \rho
J_{\frac{|m|}{n}} \Big(\frac{\nu \beta_n}{2t} \rho \Big)\Bigg]_{n=1}.
\label{12}
\end{eqnarray}
                                                                                                                                                                                               
In obtaining the last expression we have used the identities:
\begin{eqnarray}
 \sum_m i^{-m} J_m (z)=\textrm{e}^{-iz},\ \sum_m J_m^2 (z)=1. \label{13}
\end{eqnarray}
                                                                                              
In Appendix A we prove that:
\begin{eqnarray}
\frac{1}{2it} \sum_m \mathop \int \limits_0^\infty d\rho \rho \textrm{e}^{-\frac{\rho^2}{2it}} i^{-\frac{|m|}{n}} J_{\frac{|m|}{n}}
\Big(\frac{\rho^2}{2t}\Big)
 \underset{it\rightarrow T}{=}&n \frac{V_2}{4\pi T}+\frac{1}{12}\Big(\frac{1}{n}-n\Big)+\mathcal{O}\Big(\frac{T}{V_2}\Big)
\label{14}
\end{eqnarray}
and
\begin{eqnarray}
\frac{1}{2it}\sum_m \mathop \int \limits_0^{\infty} d\rho \rho J^2_{\frac{|m|}{n}}\Big(\frac{\nu \beta_n}{2t}\rho\Big)
\underset{it\rightarrow T}{=} \frac{V_2}{4\pi T}+\mathcal{O}\Big(\frac{T}{V_2}\Big). \label{15}
\end{eqnarray}
In the above equations we introduced an upper cutoff $R$ in the $\rho$-integrals and we have written as $V_2=\pi R^2$ the volume of
the two dimensional subspace. Substituting the first term in the rhs of (\ref{14}) into (\ref{12}) and feeding with the result 
(\ref{12}) we find (see Appendix) that it leads to the logarithm of the partition function:
\begin{eqnarray}
\frac{1}{2}\frac{V_{D-1}V_2}{(4\pi)^{\frac{D+1}{2}}} \mathop \int \limits_{0^+}^{\infty} \frac{dT}{T^{\frac{D+3}{2}}} \textrm{e}^{-Tm^2}
\sum_{\nu} \textrm{e}^{-\frac{(\nu \beta)^2}{4T}}=\ln Z_1 (\beta). \label{16}
\end{eqnarray}
Following the same steps for the first term in the rhs of (\ref{15}) we can prove that it is connected to the thermal entropy of the system:
\begin{eqnarray}\nonumber
 &&\frac{1}{2}\frac{V_{D-1}V_2}{(4\pi)^{\frac{D+1}{2}}} \partial_n \Bigg[\mathop \int \limits_{0^+}^{\infty} \frac{dT}{T^{\frac{D+3}{2}}} \textrm{e}^{-Tm^2}
\sum_{\nu} \textrm{e}^{-\frac{(\nu \beta)^2}{4T}}\Bigg]\\
 &=&\frac{1}{2}V_D \int \frac{d^D p}{(2 \pi)^D}\Bigg(\ln(1-\textrm{e}^{-\beta\omega})
-\frac{\beta\omega}{\textrm{e}^{\beta\omega}-1}\Bigg)
\label{17}
\end{eqnarray}
where $\omega^2=p^2+m^2$.
The contribution to (\ref{12}) of the second term in the rhs of (\ref{14}) assumes the form:
\begin{eqnarray}\nonumber
 &&\frac{1}{12}\frac{1}{(4\pi)^{\frac{D-1}{2}}}V_{D-1} \mathop \int \limits_0^{\infty} \frac{dT}{T^{\frac{D+1}{2}}} \textrm{e}^{-Tm^2}
\sum_{\nu} \textrm{e}^{-\frac{(\nu\beta)^2}{4T}}\\
 &=&\frac{\pi}{6} V_{D-1} \int \frac{d^Dp}{(2\pi)^D\omega}\frac{1}{\tanh (\frac{\omega\beta}{2})}.
\label{18}
\end{eqnarray}
Collecting everything together and using (\ref{5}) we get the geometric entropy:
\begin{eqnarray}\nonumber
 S_g&=&\frac{\pi}{6} V_{D-1} \int \frac{d^Dp}{(2\pi)^D\omega}\frac{1}{\tanh (\frac{\omega\beta}{2})}+\\
 &&+\frac{1}{2}V_D \int \frac{d^D p}{(2 \pi)^D}\Bigg(\ln(1-\textrm{e}^{-\beta\omega})-\frac{\beta\omega}{\textrm{e}^{\beta\omega}-1}\Bigg).
\label{19}
\end{eqnarray}
At the limit $p\rightarrow \infty$ , $\tanh\Big(\frac{\sqrt{p^2+m^2}\beta}{2}\Big)\rightarrow 1$ and consequently the fist integral in 
(\ref{19}) diverges. The same divergence appears in the case of zero temperature: 
\begin{eqnarray}\nonumber
 S_g(\beta=\infty)&=&\frac{\pi}{6}V_{D-1} \int \frac{d^Dp}{(2\pi)^D}\frac{1}{\sqrt{p^2+m^2}}\\
 &\rightarrow& \frac{1}{12} \frac{V_{D-1}}{(4\pi)^{\frac{D-1}{2}}} m^{D-1} \Gamma \Big(-\frac{D-1}{2},\frac{m^2}{\Lambda^2}\Big).
\label{20}
\end{eqnarray}
After this observation we are led to write:
\begin{eqnarray}
 S_g(\beta)&=&S_g(\beta=\infty)+
\frac{\pi}{3}V_{D-1} \int \frac{d^Dp}{(2\pi)^D\omega} \frac{1}{\textrm{e}^{\beta\omega}-1}+\frac{1}{2}S_{\rm thermal}.
\label{21}
\end{eqnarray}                                                  
Some comments are in order at this point. The first term in the last expression represents the well known \cite{Kabat94,Callan94,Calabrese04}
entanglement entropy at zero temperature. This is a quantity that diverges in the absence of an ultraviolet cutoff,
while it grows like the boundary surface of the excluded subregion. This fact clearly indicates the existence of very
strong quantum correlations between fields defined at neighboring points, a direct consequence of a local quantum field theory.
A quantitative explanation of such a behavior can be traced back to the uncertainty relations. Even at zero temperature,
the notion of a sharp, well defined, boundary surface is more classical than quantum.  The divergences appearing in $S_g(\textrm{T}=0)$ are connected
to the fact that in (\ref{20}) we integrate down to zero distance, driving to infinity the density of the reduced density matrix eigenvalues.
The second term is finite and well-defined for $m^2>0$.  It is also proportional to the boundary surface and it is an increasing function
of the temperature. We can consider it as a measure of the number of degrees of freedom that have been excited on the boundary surface
due to the non-zero temperature and, consequently, as a measure of the thermal correlations between the partitions. 
The last term is the thermal entropy of the subsystem, an obviously extensive quantity.
Subtracting this term from the geometric entropy we are led to define the mutual information:
\begin{eqnarray}
I_m(\beta)=S_g(\beta=\infty)+\frac{\pi}{3}V_{D-1} \int \frac{d^Dp}{(2\pi)^D \omega} \frac{1}{\textrm{e}^{\beta \omega}-1}. \label{22}
\end{eqnarray}
In general the mutual information is a measure of all correlations, thermal and quantum. We use the following definition:
\begin{eqnarray}
I(A:B)=S(\rho_A)+S(\rho_B)-S(\rho_{AB}). \label{mutual}
\end{eqnarray}
For the case in hand the entropy of the combined system $AB$ is just the total thermal entropy,
$S(\rho_{AB})=S_{\rm thermal}$. 
The entropies of each one of the two subsystems are equal due to the way we have divided our system. Moreover, each one of 
them contains a part which is one half of the total thermal entropy of the system. Thus, their  extensive thermal contribution to the mutual information
is equal to $S(\rho_{AB})$ and when substracting the latter, all contributions due to the thermal entropy will be eliminated.
So, what (\ref{22}) represents is the mutual information of the system divided by 2,
$I_m(\beta)\equiv I(A:B)/2.$


\section{Mutual information and Bose-Einstein condensation}
Almost all of the technical details needed for the current section have already been exposed in the previous one.
The basic difference of the analysis that follows, lies on the fact that we are now interested in charged scalar
(non- interacting) fields. The field theoretical description will be based on complex fields while the introduction of a
chemical potential (as a Lagrange multiplier) 
will ensure the conservation of the charge. In this framework, the partition function of the system assumes the form:
\begin{eqnarray}
Z(\beta)=\mathop {\int \limits_{\beta-\rm{periodic}}\mathcal{D}\phi \mathcal{D}\phi^*}
\exp\Bigg\{-\mathop \int \limits_0^\beta d\tau \int d^D x \mathcal{L}[\phi,\phi^*]\Bigg\}. \label{23}
\end{eqnarray}
The Lagrangian entering the last expression can be written \cite{Kapusta81,Haber81} as follows:
\begin{eqnarray}
\mathcal{L}[\phi,\phi^*]=\phi^*\big[-(\partial_{\tau}-\mu)^2-\partial_{\vec{x}}^2+m^2\big]\phi. \label{24}
\end{eqnarray}
Following the same  steps as in the previous section, we find:
\begin{eqnarray}
\ln Z_n(\beta)=\frac{1}{(4\pi)^{\frac{D-1}{2}}}V_{D-1}\mathop \int \limits_0^{\infty} \frac{dT}{T^{\frac{D+1}{2}}} \textrm{e}^{-Tm^2}
\textrm{tr}_{C_n} \textrm{e}^{-T(-\partial_E^2+2\mu\partial_0-\mu^2)}.
\label{25}
\end{eqnarray}
Once again we start from the free thermal propagator in Cartesian coordinates:
\begin{eqnarray}\nonumber
 A_{\beta_n}(\vec{x}',\vec{x})&=&\frac{1}{4\pi T} \sum_{\nu=-\infty}^{\infty} \exp \Bigg\{-\frac{1}{4T}(\vec{x}'-\vec{x})^2+\\
 &&+\Big(\frac{\nu\beta_n}{2T}+\mu\Big)(x_0'-x_0)-\frac{(\nu \beta_n)^2}{4T}-\mu\nu\beta_n \Bigg\}
\label{26}
\end{eqnarray}
to arrive at the traced quantity that is relevant for the final calculation in (\ref{25}):
\begin{eqnarray}\nonumber
 \textrm{tr}_{C_n} \textrm{e}^{-T(-\partial_E^2+2\mu\partial_0-\mu^2)}
&\rightarrow& n \textrm{tr}_{C_1} \textrm{e}^{-T(-\partial_E^2+2\mu\partial_0-\mu^2)}+\\
&&+\frac{1}{12}\Bigg(\frac{1}{n}-n\Bigg)
\sum_\nu \textrm{e}^{-\frac{(\nu\beta)^2}{4T}-\mu\nu\beta}+\mathcal{O}\Big(\frac{T}{V_2}\Big)
\label{27}
\end{eqnarray}
where the arrow underlines the fact that we have kept only the terms that are relevant for determining the mutual information.

In Appendix A we show that:
\begin{eqnarray}
 I_m&=&\frac{\pi}{6}V_{D-1}\int \frac{d^Dp}{(2\pi)^D\omega} \Bigg\{\frac{1}{\tanh\Big[\frac{(\omega-\mu)\beta}{2}\Big]}+\frac{1}{\tanh\Big[\frac{(\omega+\mu)\beta}{2}\Big]}\Bigg\}.
\label{28}
\end{eqnarray}
Isolating the (diverging) zero temperature contribution we find:
\begin{eqnarray}\nonumber
 I_m&=&S_g(\beta=\infty)+\frac{\pi}{6}V_{D-1}
 \int \frac{d^Dp}{(2\pi)^D\omega} \Bigg\{\frac{1}{\textrm{e}^{(\omega-\mu)\beta}-1}+\\
 &&+\frac{1}{\textrm{e}^{(\omega+\mu)\beta}-1}\Bigg\}.
\label{29}
\end{eqnarray}
For the system in hand the zero temperature entanglement entropy reads:
\begin{eqnarray}\nonumber
 S_g(\beta=\infty)&=&\frac{\pi}{3}V_{D-1} \int \frac{d^Dp}{(2\pi)^D\omega}\\
 &\rightarrow& \frac{1}{6} \frac{V_{D-1}}{(4\pi)^{\frac{D-1}{2}}}m^{D-1} \Gamma\Big(-\frac{D-1}{2},\frac{m^2}{\Lambda^2}\Big).
\label{30}
\end{eqnarray}

To reveal the physical content of our results we shall focus on the well-studied $D=3$ case which hosts the Bose-Einstein condensation.
As it is well-known \cite{Landsberg65,Kapusta81,Haber81} the quantitative realization of the phenomenon is different at the two opposite limits,
the non-relativistic $\rho\ll m^3$ and the ultra-relativistic one $\rho\gg m^3$, as these are defined by the total charge density of the system.

Beginning from the non-relativistic case, in which the charge density is very low and the anti-particle contribution can be omitted \cite{Haber81},
we rewrite (\ref{29}) in the form:
\begin{eqnarray}
I_m^{NR}(\beta)&=&S_g(\beta=\infty)+\frac{\pi}{6}\frac{V_2}{m}\int\frac{d^3p}{(2\pi)^3}\frac{1}{\textrm{e}^{(\frac{p^2}{2m}-\mu_{NR})\beta}-1}
\label{31}
\end{eqnarray}
where we noted as $\mu_{NR}(\beta)=\mu-m\leq 0$ the non-relativistic chemical potential. 
The integral appearing in (\ref{31}) is the total density of particles occupying excited states:
\begin{eqnarray}
  \rho^e&=&\int \frac{d^3p}{(2\pi)^3} \frac{1}{\textrm{e}^{(\frac{p^2}{2m}-\mu_{NR})\beta}-1}
  =\Bigg(\frac{m}{2\pi\beta}\Bigg)^{3/2} \sum_{n=1}^{\infty} \frac{z_{NR}^n}{n^{3/2}},
\label{32}
\end{eqnarray}
where $\ z_{NR}=\textrm{e}^{\beta\mu_{NR}}\leq 1$.
For temperatures below a certain critical value $T_C$ we have $\mu(T_C)=m$ and the above quantity is a constant:
\begin{eqnarray}
 \rho^e &=&\Bigg(\frac{m}{2\pi\beta}\Bigg)^{3/2}\sum_{n=1}^{\infty}\frac{1}{n^{3/2}}
 =\Bigg(\frac{2\pi m}{\beta}\Bigg)^{3/2}\zeta\Big(\frac{3}{2}\Big),\ 0\leq \textrm{T}\leq \textrm{T}_C.
\label{33}
\end{eqnarray}

At exactly the critical temperature the number (\ref{33}) becomes the conserved total particle density of the system:
\begin{eqnarray}
\rho^e=\rho=\Bigg(\frac{2\pi m}{\beta_C}\Bigg)^{3/2}\zeta\Big(\frac{3}{2}\Big). \label{34}
\end{eqnarray}
As an immediate consequence we get for the mutual information:
\begin{eqnarray}
I^{NR}_{m}=S_g(\textrm{T}=0)+\frac{\pi}{6}\frac{V_2}{m}\rho\Bigg(\frac{\textrm{T}}{\textrm{T}_C}\Bigg)^{3/2},\ \textrm{T}<\textrm{T}_C. \label{35}
\end{eqnarray}

Above the critical temperature the system passes to the gas phase in which all of the particles occupy excited states. The mutual information reads now:
\begin{eqnarray}
I^{NR}_{m}=S_g(\textrm{T}=0)+\frac{\pi}{6}\frac{V_2}{m}\rho,\ \textrm{T}>\textrm{T}_C. \label{36}
\end{eqnarray}
Thus, the Bose-Einstein condensation and the relevant phase transition are reflected in a discontinuity of the derivative of the mutual information:                                  
\begin{eqnarray}
\frac{\partial I_m^{NR}}{\partial \textrm{T}} \Bigg|_{\textrm{T}=\textrm{T}_C^-}-\frac{\partial I_m^{NR}}{\partial \textrm{T}} \Bigg|_{\textrm{T}=
\textrm{T}_C^+} =\frac{\pi^2}{2}\zeta^{2/3}\Big(\frac{3}{2}\Big) V_2 \rho^{1/3}. \label{37}
\end{eqnarray}

When $\rho\gg m^3$ we are approaching the ultra-relativistic limit, the critical temperature rises at relativistic high values
$\textrm{T}_C=(3|\rho|/m)^{1/2} \gg m$ and the behavior of the system changes.
Below the critical temperature, one easily finds that \cite{Landsberg65,Kapusta81}: 
\begin{eqnarray}\nonumber
 I_m&=&S_g(\textrm{T}=0)+\frac{\pi}{6}V_2 \int \frac{d^3p}{(2\pi)^3\omega}\Bigg(\frac{1}{\textrm{e}^{(\omega-\mu)\beta}-1}+\frac{1}{\textrm{e}^{(\omega+\mu)\beta}-1}\Bigg)\\
&\approx& S_g(\textrm{T}=0)+\frac{\pi V_2}{12}\frac{|\rho|}{m}\Bigg(\frac{\textrm{T}}{\textrm{T}_C}\Bigg)^2,\ \textrm{T}<\textrm{T}_C.
\label{38}
\end{eqnarray}
The integral that appears in (\ref{37}) and (\ref{38}) is not the charge density of the system and, consequently, is not a conserved quantity even for
temperatures above the critical one.  However, it is not hard to confirm \cite{Landsberg65} that at high temperatures $\textrm{T}>\textrm{T}_C$
it behaves as following:
\begin{eqnarray}\nonumber
 &&\int\frac{d^3p}{(2\pi^3)\omega}\Bigg(\frac{1}{\textrm{e}^{(\omega-\mu)\beta}-1}+\frac{1}{\textrm{e}^{(\omega+\mu)\beta}-1}\Bigg)\\
\nonumber&=&\frac{\textrm{T}^2}{6}-\frac{\textrm{T}}{2\pi}(m^2-\mu^2)^{1/2}-\frac{m^2}{4\pi^2}\ln\Big(C\frac{m}{\textrm{T}}\Big)+\\
&&+\frac{1}{4\pi^2}(m^2-\mu^2)+\mathcal{O}\Big(\frac{m^2}{\textrm{T}^2}\Big)
\label{39}
\end{eqnarray}
where $C=\textrm{e}^{\gamma_E-1}/4\pi$ and $\gamma_E$ is the Euler-Macheroni constant.
 
As in the non-relativistic case, the derivative of the mutual information with respect to the temperature possesses a discontinuity that
reflects the underlying phase transition:
\begin{eqnarray}
\frac{\partial I_m}{\partial \textrm{T}} \Bigg|_{\textrm{T}=\textrm{T}_C^-}-\frac{\partial I_m}{\partial \textrm{T}} \Bigg|_{\textrm{T}=\textrm{T}_C^+}
=-\frac{\pi \sqrt{3}}{9} V_2 \Bigg(\frac{|\rho|}{m}\Bigg)^{1/2}.
\end{eqnarray}
The last result completes our study for the influence of the Bose-Einstein condensation on the entropy of entanglement in an ideal Bose system at
finite temperature and non-zero chemical potential.

\section{Conclusion}
In the current work we have performed two types of calculations and we have arrived at results with a clear physical content.
First, we calculated the geometric entropy in an ideal Bose system at finite temperature and we confirmed the expected result:
It combines the genuine quantum correlations with the  thermal fluctuations, and it becomes an extensive quantity for an
infinite system. Due to the simplicity of the system under consideration, we were able to explicitly subtract from
the geometric entropy its extensive component which coincides with the corresponding thermal entropy. In this way we defined the so-called mutual information, which grows like the surface that bounds the space region in which a system lives.
The second calculation we performed refers to a Bose system at finite temperature and non-zero chemical potential.
We found that, at the critical temperature, the temperature derivative of the mutual information exhibits a
finite discontinuity, and we explicitly calculated it. This result connects the condensation that appears in an ideal quantum 
Bose system with the spatial correlations between two regions of the system. This connection was shown by using a purely informational
tool namely, the quantum mutual information.

\section*{Acknowledgements}
C. N. G. acknowledges the financial support of the Wallonie-Bruxelles International doctoral fellowship of excellence.

\appendix
\section*{Appendix}
\setcounter{section}{1}
In this Appendix we shall prove those of the formulas appearing in the text for which summations over $m$ or $\nu$ must be performed.
To begin with, let us discuss (\ref{14}). The relevant integral diverges and calls for the introduction of a cutoff:
\begin{eqnarray}
\frac{1}{2it}\sum_m\mathop\int\limits_0^{\infty} d\rho \rho \textrm{e}^{-\frac{\rho^2}{2it}}
i^{-\frac{|m|}{n}}J_{\frac{|m|}{n}}\Big(\frac{\rho^2}{2t}\Big)\underset{it\rightarrow T}{\rightarrow}&\frac{1}{2T}\sum_m\mathop\int_0^R d\rho\rho\textrm{e}^{-\frac{\rho^2}{2T}}I_{\frac{|m|}{n}}\Big(\frac{\rho^2}{2T}\Big).
\label{A.1}
\end{eqnarray}
To handle the last integral we make an intermediate step by introducing the following expression:
\begin{eqnarray}\nonumber
F_n(\alpha)&=&\frac{1}{2}\sum_{m=-\infty}^{\infty}\mathop\int\limits_0^\infty dq \textrm{e}^{-\alpha q}I_{\frac{|m|}{n}}(q)\\
&=&\frac{1}{2\sqrt{\alpha^2-1}} \coth \Big(\frac{\alpha+\sqrt{\alpha^2-1}}{2}\Big)
\label{A.2}
\end{eqnarray}
which is also a regularized version (for $\alpha\rightarrow 1^+$) of the integral entering (\ref{14}).
Taking the limit $\alpha=1+\epsilon,\ \epsilon\rightarrow 0^+$ one easily finds that:
\begin{eqnarray}
F_n(\alpha)=\frac{n}{2\epsilon}+\frac{1}{12}\Big(\frac{1}{n}-n\Big)+\mathcal{O}(\epsilon). \label{A.3}
\end{eqnarray}
We immediately see that the diverged part of the integral appears for $n=1$.
In this case the integration in (\ref{A.1}) is trivial and we are led to the conclusion:
\begin{eqnarray}
\frac{1}{\epsilon}\rightarrow\frac{\pi R^2}{4\pi T}=\frac{V_2}{4\pi T}. \label{A.4}
\end{eqnarray}
Combining this identification with the finite part appearing in (\ref{A.3}) we get the confirmation of (\ref{14}). 

Our next concern is (\ref{16}). Using the identities:
\begin{eqnarray}\nonumber
\textrm{e}^{-\frac{(\nu\beta)^2}{4T}}&=&(4\pi T)^{\frac{D+1}{2}} \int \frac{d^{D+1}p}{(2\pi)^{D+1}}\textrm{e}^{-Tp^2+ip_0\nu\beta},\\
\sum_{\nu=-\infty}^\infty \textrm{e}^{ip_0\nu\beta}
&=&\frac{2\pi}{\beta}\sum_{k=-\infty}^{\infty}\delta(p_0-\omega_k),\ \omega_k=\frac{2\pi k}{\beta}
\label{A.5}
\end{eqnarray}
we recast the integral appearing in Eq. (\ref{16}) into the form:
\begin{eqnarray}\nonumber
&& \frac{1}{2}\frac{V_{D-1}V_2}{(4\pi)^{\frac{D+1}{2}}} \mathop \int \limits_{0^+}^{\infty} \frac{dT}{T^{\frac{D+3}{2}}}
\textrm{e}^{-Tm^2}\sum_{\nu} \textrm{e}^{-\frac{(\nu\beta)^2}{4T}}\\
&=&\frac{V_D}{4}\int\frac{d^Dp}{(2\pi)^D}\sum_k \mathop \int \limits_{0^+}^{\infty}\frac{dT}{T} \textrm{e}^{-T\big[(\beta\omega)^2+(2\pi k)^2\big]}.
\label{A.6}
\end{eqnarray}
To obtain the last result we wrote $V_2=\mathop \int \limits_0^{\beta} d\tau \mathop \int \limits_{-\infty}^{\infty} dx_1 \rightarrow \beta L$,
we rescaled $T \rightarrow T \beta^2$ and we used the abbreviation $\omega^2=p^2+m^2$. 

Performing the integral over $T$ and neglecting an irrelevant (infinite) constant we get:
\begin{eqnarray}
\mathop \int \limits_{0^+}^{\infty}\frac{dT}{T} \textrm{e}^{-T\big[(\beta\omega)^2+(2\pi k)^2\big]}
=-\ln\big((\beta\omega)^2+(2\pi k)^2\big). \label{A.7}
\end{eqnarray}
The summation over $k$ is standard \cite{Kapusta81}:
\begin{eqnarray}
\frac{1}{2}\sum_k\ln\big((\beta\omega)^2+(2\pi k)^2\big)=\frac{1}{2}\beta\omega+\ln(1-\textrm{e}^{-\beta \omega}). \label{A.8}
\end{eqnarray}
The last result proves (\ref{16}).

Following the same line of reasoning we can prove (\ref{18}). Using, once again, the identities (\ref{A.5}) we rewrite the relevant integral in the form:
\begin{eqnarray}\nonumber
 && \frac{1}{12}\frac{V_{D-1}}{(4\pi)^{\frac{D-1}{2}}} \mathop \int \limits_{0}^{\infty} \frac{dT}{T^{\frac{D+1}{2}}} \textrm{e}^{-Tm^2}
\sum_{\nu} \textrm{e}^{-\frac{(\nu\beta)^2}{4T}}\\
 &=&\frac{\pi}{3\beta}V_{D-1} \int \frac{d^Dp}{(2\pi)^D}\sum_k \frac{1}{\omega^2+\omega_k^2}.
\label{A.9}
\end{eqnarray}
The summation is easily performed:
\begin{eqnarray}
\sum_k \frac{1}{\omega^2+\omega_k^2}=\frac{\beta}{2\omega}\frac{1}{\tanh(\beta\omega)}. \label{A.10}
\end{eqnarray}
Combining (\ref{A.9}) and (\ref{A.10}) we immediately obtain (\ref{18}) of the text.

Our next concern is Eq. (\ref{15}). The relevant integral diverges and the introduction of a cutoff is necessary. To this end let us discuss the integral: 
\begin{eqnarray}
\sum_m \mathop \int \limits_0^{\infty} d\rho\rho \textrm{e}^{-\frac{\rho^2}{R^2}} J^2_{\frac{|m|}{n}}\Big(\frac{\nu\beta_n}{2t}\rho\Big)
=\frac{R^2}{2}\textrm{e}^{-\frac{(\nu \beta R)^2}{8t^2}}\sum_m I_{\frac{|m|}{n}}\Big(\frac{(\nu\beta R)^2}{8t^2}\Big)=\frac{R^2}{2}
\label{A.11}
\end{eqnarray}
which can be considered (at the limit $R\rightarrow \infty$) as a regularized version of the integral appearing in (\ref{15}).
Note that the divergence in (\ref{A.11}) is independent of $n$ and, contrary to (\ref{A.3}), there is no finite part for $n\neq 1$.
This completes the proof of (\ref{15}).

To prove (17) it is enough to follow the road we followed to arrive at (\ref{A.8}). Beginning from the relation:
\begin{eqnarray}\nonumber
 &&\frac{1}{2}\frac{V_{D-1}V_2}{(4\pi)^{\frac{D+1}{2}}} \mathop \int \limits_{0^+}^{\infty} \frac{dT}{T^{\frac{D+3}{2}}}
\textrm{e}^{-Tm^2}\sum_{\nu} \textrm{e}^{-\frac{(\nu\beta_n)^2}{4T}}\\
 &=&\frac{V_D}{4n}\int\frac{d^Dp}{(2\pi)^D}\sum_k \mathop \int \limits_{0^+}^{\infty}\frac{dT}{T} \textrm{e}^{-T[(\beta_n\omega)^2+(2\pi k)^2]}
\label{A.12}
\end{eqnarray}
we only have to perform a differentiation with respect to $n$ to arrive at the result indicated in (\ref{17}):
\begin{eqnarray}\nonumber
 && \partial_n\Bigg\{\frac{1}{n}\int \frac{d^Dp}{(2\pi)^D}
\Big[-\frac{1}{2}\beta\omega n -\ln \Big(1-\textrm{e}^{-\beta\omega n}\Big)\Big]\Bigg\}\Bigg|_{n=1}\\
 &=&\int\frac{d^Dp}{(2\pi)^D} \Big[\ln \Big(1-\textrm{e}^{-\beta\omega}\Big)-\frac{\beta \omega}{\textrm{e}^{\beta\omega}-1}\Big].
\label{A.13}
\end{eqnarray}

The last relation we have to prove is (\ref{28}) of the text. We begin by using the Poisson summation formula to find that:
\begin{eqnarray}\nonumber
  \sum_{\nu} \textrm{e}^{-\frac{(\nu\beta)^2}{4T}-\mu\nu\beta}&=&\sum_k \mathop \int \limits_{-\infty}^{\infty} dx
\textrm{e}^{2\pi i k x}\textrm{e}^{-\frac{(x\beta)^2}{4T}-\mu x \beta}\\
&=&\frac{\sqrt{4\pi T}}{\beta}\sum_k \textrm{e}^{-T (\omega_k+i\mu)^2}.
\label{A.14}
\end{eqnarray}
With the help of this result we get the mutual information:
\begin{eqnarray}
I_m=\frac{2\pi}{3\beta}V_{D-1}\int\frac{d^Dp}{(2\pi)^D} \sum_k \frac{1}{(\omega_k+i\mu)^2+\omega^2}. \label{A.15}
\end{eqnarray}
The sum in the last expression can be easily performed if we rewrite it in the form:
\begin{eqnarray}\nonumber
 && \sum_k \frac{1}{(\omega_k+i\mu)^2+\omega^2}\\
&=& \frac{1}{2}\frac{\omega-\mu}{\omega}\sum_k \frac{1}{\omega_k^2+(\omega-\mu)^2}
+\frac{1}{2}\frac{\omega+\mu}{\omega}\sum_k \frac{1}{\omega_k^2+(\omega+\mu)^2}.
\label{A.16}
\end{eqnarray}
Using for each term the formula (\ref{A.10}) we get the result indicated in (\ref{28}).

It would be useful to compare our result indicated in (\ref{19}) with the corresponding result derived in the framework 
of a two dimensional conformal scalar field theory \cite{Calabrese04} with central charge $c=1/2$.
This can be done by identifying the ultraviolet cutoff $\Lambda$ with the inverse lattice spacing $1/\alpha$ and the mass $m$
with the inverse finite size of excluded interval $1/l$ (that is, the infrared cutoff). 
Given that our result is valid at the limit $L^2m^2\rightarrow \infty$, the comparison is meaningful only for
$\beta/\alpha\rightarrow \infty$ or $l/\beta\rightarrow\infty$.
Applying (\ref{19}) for $D=1$ we get the result:
\begin{eqnarray}
S_g = \left\{ \begin{array}{ll}
\frac{1}{6}\ln\Big(\frac{\Lambda}{m}\Big)\rightarrow\frac{1}{6}\ln \Big(\frac{l}{\alpha}\Big) & \textrm{if $\beta\rightarrow\infty$}\\
\frac{\pi}{6}\frac{1}{m\beta}\rightarrow\frac{\pi}{6}\frac{l}{\beta} & \textrm{if $\beta\rightarrow 0$}
\end{array} \right.. \label{A.17}
\end{eqnarray}


\end{document}